%% file: BScausality.tex
\documentclass[12pt]{article}

\usepackage{graphicx}
\usepackage{amssymb}
\usepackage{amsmath}
\usepackage{srcltx}
\usepackage{color}
\usepackage{cite}
\usepackage{url}
\usepackage{ntheorem}
\usepackage{psfrag}
\usepackage[bookmarks,colorlinks=false]{hyperref}

\theorembodyfont{\upshape}

\theoremsymbol{\ensuremath{\diamondsuit}}
\newcommand{\fcoco}{\small}
\theorembodyfont{\fcoco}\theoremseparator{}\theoremindent0.5cm

\theoremstyle{nonumberplain}\theorembodyfont{\fcoco}
\theoremseparator{}\theoremindent0.5cm

\DeclareFontFamily{OT1}{rsfs}{}
\DeclareFontShape{OT1}{rsfs}{m}{n}{ <-7> rsfs5 <7-10> rsfs7 <10-> rsfs10}{}
\DeclareMathAlphabet{\mycal}{OT1}{rsfs}{m}{n}

{\catcode `\@=11 \global\let\AddToReset=\@addtoreset}
\AddToReset{equation}{section}

\newcounter{mnotecount}[section]

\AddToReset{figure}{section}
\renewcommand{\themnotecount}{\thesection.\arabic{mnotecount}}

\newcommand{\mnotex}[1]
{\protect{\stepcounter{mnotecount}}$^{\mbox{\footnotesize
$
\bullet$\themnotecount}}$ \marginpar{
\raggedright\tiny\em
$\!\!\!\!\!\!\,\bullet$\themnotecount: #1} }

\input{LeeNewcommand}

\begin{document}
\title{Stable causality of Black Saturns}

\author{Sebastian J. Szybka \\
Astronomical Observatory, Jagellonian University,
Krak\'ow}
\date{}

\maketitle{}
\begin{abstract}
We prove that the Black Saturns are stably causal on the closure of the domain of outer communications. 
\end{abstract}

\section{Introduction}
\label{sec:intro}
The Black Saturn solution of Elvang and Figueras is a five dimensional black hole with disjoint event horizons with topology $\R\times S^3$ and
$\R\times S^1\times S^2$ \cite{EF,CES}. It describes a spherical black hole\footnote{The adjective ``spherical'' refers to the topology of the horizon.} surrounded by a black ring.

In this article, we address the question of causality
violations in the Black Saturn solution.
The absence of closed causal curves is one of the desired
properties of a solution to the Einstein equations. Such
property should be stable against small perturbations.
Therefore, the closed causal curves are not allowed in any
Lorentzian metric that is sufficiently near the original one.
This leads to the
 notion of {\it stable causality} as
introduced by Hawking \cite{HawSC}. Technically, a spacetime
$(M,g_{\mu\nu})$ is {\it stably causal} if and only if there
exists a differentiable function $f$ on $M$ such that
$\nabla^\mu f$ is a past directed timelike vector field
\cite{HawSC,HE}. The non-unique function $f$ may be interpreted
as a cosmic time that increases along every causal curve.

The natural guess for the Black Saturn is that the generalized Weyl
$t$ coordinate is a cosmic time in the domain of outer
communications (d.o.c.). Under this hypothesis, the problem of
{\it stable causality} of the d.o.c.\ was reduced in \cite{CES}
to the study of the $g_{\psi\psi}$ component of the metric.
Namely, stable causality will result from the following
inequality in the generalized Weyl coordinates
\begin{equation}\label{eq:sc}
 g(\nabla t, \nabla t) = g^{tt} = - \frac{g_{\psi\psi}}{G_y}<0\;,
\end{equation}
where $G_y$ is a non-negative function which is zero only on the axis of the rotation of the Killing field $\partial_\psi$. In other words, it is sufficient to show that $g_{\psi\psi}$ vanishes as fast as $G_y$ on the axis of the rotation of $\partial_\psi$ and that $g_{\psi\psi}> 0$ in the remaining part of the d.o.c.

Numerical evidence for positivity of $g_{\psi\psi}$ \emph{in
the plane of the ring} (as defined in \cite[Section 3.8]{EF})
was already signalled in the original paper of Elvang and
Figueras~\cite{EF}. This numerical evidence was extended in
\cite{CES} to the part of spacetime covered by the generalized
Weyl coordinates away from points where $\partial_\psi$
vanishes. Before this work, all analytical proofs of {\it stable
causality} were restricted to the situation when the Komar
angular momentum of the spherical component of the horizon is
equal to zero: under this restriction, the positivity of
$g_{\psi\psi}$ in the plane of the ring was established in
\cite{EF}, and {\it stable causality} of the d.o.c.\ was shown
in \cite{CES}.

In this article, we prove that the d.o.c.\ of the general
Black Saturn solution is {\it stably causal}. In addition, we show that the event horizons may be included to the domain of {\it stable causality}, hence the Black Saturns are {\it stably causal} on the closure of the d.o.c.\footnote{In fact, 
our proof remains valid for a family of the Black Saturn solutions with conical singularities on the axes
of the periodic Killing fields $\partial_\psi$, $\partial_\varphi$.}

The calculations presented here\footnote{The {\sc Mathematica}
code is available at
\url{http://th.if.uj.edu.pl/~szybka/BScausality}} involve
manipulations of huge algebraic expressions and were done with
{\sc Mathematica}.
We use the same notation and definitions of auxiliary functions as in \cite{CES}.

\section{Stable causality}

In the generalized Weyl coordinates $(t,\psi,\varphi,\rho,z)$ the d.o.c.\ corresponds to 
$$\{\rho>0\}\cup\{\rho=0, z\not \in [a_5,a_4]\cup [a_3,a_2]\}\;,$$
where $a_{i=1,\dots,5}$ are parameters.
Since $g_{\psi\psi}=g_{\psi\psi}(\rho,z,a_1,\dots,a_5)$, then we would like to prove {\it stable causality} for a union of the following sets in $\R^7$
\pagebreak
\begin{eqnarray}\label{eq:sets}
\tilde V_a&=&\{a_1<a_5<a_4<a_3<a_2\}\;,\\\nonumber
\tilde V_I&=&\{\rho>0\}\cap \tilde V_a\;,\\\nonumber
\tilde V_{II}&=&\{\rho=0,z<a_1\}\cap \tilde V_a\;,\\\nonumber
\tilde V_{III}&=&\{\rho=0,a_1\leq z<a_5\}\cap \tilde V_a\;,\\\nonumber
\tilde V_{IV}&=&\{\rho=0,a_4<z<a_3\}\cap \tilde V_a\;,\\\nonumber
\tilde V_{\psi}&=&\{\rho=0,a_2<z\}\cap \tilde V_a\;.
\end{eqnarray}
Hence, we have $\tilde V_{d.o.c.}=\tilde V_I\cup \tilde
V_{II}\cup \tilde V_{III}\cup\tilde V_{IV}\cup\tilde V_{\psi}$.
We are interested in non-degenerate solutions, so the
parameters are restricted to $\tilde V_a$. The ``plane of the
ring'' corresponds to $\tilde V_{II}\cup \tilde V_{III}\cup
\tilde V_{IV}$, while $\tilde V_\psi$ is the intersection of the rotation axis of
$\partial_\psi$ with  the d.o.c. 
The event horizons of the black ring and the spherical
component coincide with $(\{\rho=0\}\setminus\tilde V_{d.o.c.})\cap \tilde V_a$. We would like to show that $g_{\psi\psi}>
0$ on $\tilde V_{I}\cup \tilde V_{II}\cup \tilde V_{III}\cup
\tilde V_{IV}$. Moreover, since $G_y$ vanishes as $\rho^2$ on
the axis of $\partial_\psi$ \cite{CES}, then it is necessary to
check that
$$\lim_{\rho\rightarrow 0^+}\frac{g_{\psi\psi}}{\rho^2}>0$$
on $\tilde V_\psi$.

It turns out to be convenient to view $g_{\psi\psi}$ as a
function of $\rho$, $\mu_1,\dots,\mu_5$, where
$\mu_i=\sqrt{\rho^2+(z-a_i)^2}-(z-a_i)$. In this
parametrization, the translational symmetry of $z$, $a_i$ is
explicit. The analogues of the sets%
\footnote{These sets are not
equivalent to \eqref{eq:sets} because $V_4\subset V_{IV}$,
$V_2\subset V_\psi$, where $V_4$, $V_2$ correspond to the sets
that are defined in the old parametrization as
$\{\rho=0,z=a_4\}\cap \tilde V_a$, $\{\rho=0,z=a_2\}\cap \tilde
V_a$, respectively.}
\eqref{eq:sets} are now in $\R^6$
\begin{eqnarray}\label{eq:setsmu}
 V_\mu&=&\{\rho\leq\mu_1\leq\mu_5\leq\mu_4\leq\mu_3\leq\mu_2\}\;,\\\nonumber
 V_I&=&\{\rho>0\}\cap  V_\mu\;,\\\nonumber
 V_{II}&=&\{\rho=0,\mu_1>0\}\cap  V_\mu\;,\\\nonumber
 V_{III}&=&\{\rho=0,\mu_1=0,\mu_5>0\}\cap  V_\mu\;,\\\nonumber
 V_{IV}&=&\{\rho=0,\mu_4=0,\mu_3>0\}\cap  V_\mu\;,\\\nonumber
 V_{\psi}&=&\{\rho=0,\mu_2=0\}\cap  V_\mu\;,
\end{eqnarray}
and, of course, $V_{d.o.c.}\subset V_I\cup  V_{II}\cup
V_{III}\cup V_{IV}\cup V_{\psi}$. It follows from the definition of $\mu_i$ that if $\mu_i=\mu_j$ for $i\neq j$ then $\rho=\mu_i=\mu_j=0$. This parametrization turns out to be more helpful in completing the proof.

The numerator and the denominator of $g_{\psi\psi}$, when
written as polynomials in  $\rho$, $\mu_i$, and some $c_1$,
$c_2$, $q$ contain tens of thousands monomials \cite{EF,CES}.
One may check with a direct {\sc Mathematica} calculation that
some non-trivial factors from the numerator and the denominator
cancel and the original form of $g_{\psi\psi}$ may be
simplified to
\bel{eq:gPP}
g_{\psi\psi}=\frac{\mu_4\mu_5A^2-\mu_3B^2}{\mu_1\mu_4 H_x F}\;,
\ee
where
\begin{equation}\label{eq:def}
\begin{array}{l}
A=(\mu_2 p_1 ( \mu_5p_2 +c_2 q \mu_1  \mu_3p_3)+c_1  \mu_3p_4 (-q  \rho^2p_5+c_2 \mu_1 \mu_4p_6))\;,\\
B=( \mu_5p_1 (q \mu_1p_2-c_2 \mu_4 \rho^2p_3)+c_1 \mu_2 \mu_4p_4 (\mu_1p_5 \mu_5+c_2 q \mu_3p_6))\;,\\
p_1=(\mu_3-\mu_1)(\mu_1\mu_4+\rho^2)\;,\\
p_2=(\mu_2-\mu_4)(\mu_1\mu_2+\rho^2)(\mu_2\mu_3+\rho^2)\;,\\
p_3=(\mu_2-\mu_1)(\mu_2\mu_5+\rho^2)\;,\\
p_4=\mu_1(\mu_5-\mu_1)\;,\\
p_5=(\mu_2-\mu_1)(\mu_2-\mu_4)(\mu_2\mu_3+\rho^2)\;,\\
p_6=(\mu_1\mu_2+\rho^2)(\mu_2\mu_5+\rho^2)\;,
\end{array}
\end{equation}
and $p_i\geq 0$. The functions $H_x$, $F$ are non-negative and
they were defined in \cite{EF}. It follows from the analysis in
\cite{CES} that zeros of $H_x F$ exist only for $\rho=0$ and
they cancel with the zeros of the numerator of $g_{\psi\psi}$.

The parameters $c_1$, $c_2$, $q$ depend only on $a_i$ and do not depend on $\rho$, $z$. However, if one assumes that $\rho$, $\mu_i$ are independent variables then $c_1$, $c_2$, $q$ are finite continuous functions of $\rho$, $\mu_i$ and are given by
\begin{eqnarray}\label{eq:c1}
c_1^2&=&\frac{(\mu_3-\mu_1) (\mu_4-\mu_1) \mu_5 (\mu_1 \mu_3+\rho^2) (\mu_1 \mu_4+\rho^2)}{\mu_1 \mu_3 \mu_4 (\mu_5-\mu_1) (\mu_1 \mu_5+\rho^2)}\;,\\
q&=&\frac{ c_1 c_2 \mu_4(\mu_2 - \mu_1) (\mu_1 \mu_2 + \rho^2)}{c_1 \mu_1 (\mu_2 - \mu_4) (\mu_2 \mu_4 + \rho^2)+ c_2 \mu_2 (\mu_4 - \mu_1) (\mu_1 \mu_4 + \rho^2)}\label{eq:q}
\;,
\end{eqnarray}
where we imposed \eqref{eq:c1} on $c_1$ in the formula for $q$.
The formula for $c_2$ in terms of $\rho$, $\mu_i$ is to long to be usefully cited here. It may be derived from equations $(4.2)$, $(5.1)$ in \cite{CES}, but it is not necessary for our calculations.

The formulas \eqref{eq:c1}, \eqref{eq:q} are explicitly valid in $V_I\cup V_{II}$.
They are also valid in the remaining part of the d.o.c.\
provided the limit $\rho\rightarrow 0^+$ is carefully taken. If
$\mu_i(\rho=0)=0$ then taking this limit should be preceded by
the substitution $\mu_i\rightarrow\rho^2\hat{\mu}_i$, where
$\hat{\mu}_i>0$ \cite{CES}.

The simplification \eqref{eq:gPP} is a significant one. Even if
one does not substitute formulas for $c_1$, $c_2$, $q$ the
original expression for $g_{\psi\psi}$ written as a rational
function contained $106995$ monomials. This number was reduced
to $2344$ in the simplified formula.

In order to present the proof in a concise form, we introduce
the following operators. Let $N$, $D$ denote operators acting
on a rational functions that return polynomials: a numerator or
a denominator, respectively. The result is not unique and the
action of $N$, $D$ is given only up to an overall factor.
However, this non-uniqueness is not important for our problem.
We also define substitution operators $S_q$, respectively
$S_{c_1^2}$, that return the rational function which is
obtained after $q$, respectively $c_1^2$, has been substituted
in the original expression using \eqref{eq:q}, respectively
\eqref{eq:c1}.
One should note that $S_q$ and $S_{c_1^2}$ do not commute with
$N$ and $D$ in general. The formula for $c_2$, in contrast to
the formula for $c_1$, contains a square root that cannot be
eliminated by taking $c_2^2$. We prefer to preserve the
polynomial form of the evaluated expressions, hence we will
substitute only $c_1^2$ and avoid substituting $c_2$.

The calculations described below were done with {\sc
Mathematica}. We present them here in a brief form. It follows
from the smoothness of the Black Saturns \cite{CES} that the
expressions evaluated below are regular.

We start with the analysis of $g_{\psi\psi}$ on $V_I$
($\rho>0$)\footnote{$V_I=\{\rho>0\}\cap V_\mu$ as indicated in
\eqref{eq:setsmu}, but for the sake of brevity we will remind
only first part of the definitions.}. The denominator of
$g_{\psi\psi}$ is given by $\mu_1\mu_4 H_x F$ and it is a
positive function on $V_I$. The numerator of $g_{\psi\psi}$ is
equal to $\Xi_+\Xi_-$, where
\begin{equation}\label{eq:Xi}
\Xi_\pm=\sqrt{\mu_4 \mu_5}A\pm\sqrt{\mu_3}B\;.
\end{equation}
If $\Xi_+\Xi_->0$ then $g_{\psi\psi}>0$ on $V_I$, as desired.
Firstly, we check at a random point\footnote{We impose the
equation satisfied by $c_2$ in such checks, but this is
actually not necessary.} $P\in V_I$ that $\Xi_\pm|_P>0$. Since both
$\Xi_+$ and $\Xi_-$ are continuous in $\rho$, $\mu_i$, then it
is sufficient to show that they cannot vanish. The functions
$\Xi_\pm$ are linear in $q$, as may be seen from
\eqref{eq:def}, \eqref{eq:Xi}. We substitute $q$ into $\Xi_\pm$
and examine the numerators of the resulting expressions. We
would like to show that none of them ($NS_q\Xi_\pm$) has zeros.
By inspection, we find that $NS_q\Xi_\pm$ are quadratic in
$c_2$. Since $c_2$ is real, a negative discriminant of
$NS_q\Xi_\pm$ with respect to $c_2$ would imply that none of
the equations $NS_q\Xi_\pm=0$ has a solution. We calculate
these discriminants $\Delta_\pm$ and they turn out to be fourth
order in $c_1$. Next, we substitute $c_1^2$ into $\Delta_\pm$
using \eqref{eq:c1} and taking $c_1^4=(c_1^2)^2$,
$c_1^3=c_1c_1^2$. With a help of {\sc Mathematica} we have
derived
\begin{align*}
&S_{c_1^2}\Delta_\pm=w_\pm\times\nonumber\\
&\frac{\mu_1\mu_2^2 \mu_4 \mu_5^{
 5/2}(\mu_1 - \mu_2)^2 (\mu_3 - \mu_1) (\mu_4 - \mu_1) (\mu_2 - \mu_4)^2 (\mu_1 \mu_2 + \rho^2)^2 (\mu_1 \mu_4 + \rho^2)^2}{(\mu_5 - \mu_1) (\mu_1 \mu_5 + \rho^2)^2}\;.
\end{align*}
The factors multiplying $w_\pm$ are strictly positive, and
$w_\pm$ are complicated polynomials in $\rho$, $\sqrt{\mu_i}$.
These polynomials are linear in $c_1$ (with non-vanishing
coefficients in front of $c_1$ as it will follow from our
further analysis). We check at a random point $P'\in V_I$ that
$w_\pm|_{P'}<0$, hence if $w_\pm$ have no zeros then
$\Delta_\pm<0$. Let $c_1^\pm$ be solutions to the equations
$w_\pm=0$. A {\sc Mathematica} calculation reveals that
$c_1^+=-c_1^-$. We set $U=(c_1^\pm)^2-S_{c_1^2}c_1^2$ and
calculate
\begin{equation*}
U=\frac{\mu_5(\mu_1^2+\rho^2)^2}{4 \mu_1^2 \mu_3 \mu_4 (\mu_1 - \mu_5)^2 (\mu_1 \mu_5 +\rho^2)^2}\frac{\hat{U}}{\tilde{U}}\;,
\end{equation*}
where $\hat{U}$, $\tilde{U}$ are complicated
polynomials\footnote{The polynomial $\tilde U$ is a full
square.} in $\rho$, $\mu_i$ with signs unknown so far. The
coefficient in front of $\hat U/\tilde U$ is strictly positive.
Now, we succeeded in making the signs of $\hat{U}$, $\tilde{U}$
explicit by writing them in terms of the new positive functions
\begin{equation*}
 \Delta_{51}=\mu_5-\mu_1\;,\;\;\;
 \Delta_{45}=\mu_4-\mu_5\;,\;\;\;
 \Delta_{34}=\mu_3-\mu_4\;,\;\;\;
 \Delta_{23}=\mu_2-\mu_3\;.
\end{equation*}
The coefficients in $\hat{U}$, $\tilde{U}$ in front
of $\rho$, $\mu_1$, $\Delta_{ij}$,
turn out\footnote{We have $\mu_1>\rho$, but $\Delta_{ij}$ do not have
to form monotonically increasing sequence like $\mu_i$.}
to be positive and belong to
$$[9,13705432344] \cap \Z\;,\qquad [1,137075730] \cap \Z\;,$$
respectively. Since  $\rho$, $\mu_1$, $\Delta_{ij}$ are
strictly positive, then it follows that $\hat{U}$, $\tilde{U}$
are strictly positive and the equation $U=0$ does not have
solutions. Therefore, $c_1^\pm\neq S_{c_1^2}c_1$  and the
polynomials $w_\pm$ cannot vanish. This means that the
discriminants of $NS_q\Xi_\pm$ in respect to $c_2$ are negative
($\Delta_\pm<0$) and there are no real $c_2$ that would satisfy
any of the equations $\Xi_\pm=0$. Finally, this implies that
$g_{\psi\psi}>0$ on $V_I$ (for $\rho>0$), as desired. To
complete the proof it remains to repeat the analysis above in
the remaining part of the d.o.c.

For $\rho=0$ the denominator of $g_{\psi\psi}$ (given by
$\mu_1\mu_4H_x F$) is not strictly positive any more. However,
it follows from the Black Saturns' smoothness \cite{CES} that
whenever the denominator of $g_{\psi\psi}$ vanishes the
numerator of $g_{\psi\psi}$ (equal to $\Xi_+\Xi_-$) vanishes as
well and the limit $\lim_{\rho\rightarrow 0^+}g_{\psi\psi}$ is
finite. Moreover, the continuity of $g_{\psi\psi}$ implies that
this limit is non-negative, possibly zero.

On $V_{II}$ ($\rho=0$, $\mu_1>0$) the argument proceeds along
the same lines as for $V_I$. Intermediate expressions have
different form, but the reasoning is analogous. We have found
\begin{equation*}
U=\frac{\mu_1^2}{4 \mu_3 \mu_4\mu_5 (\mu_1 - \mu_5)^2}\frac{\hat{U}}{\tilde{U}}\;.
\end{equation*}
The coefficients in the polynomials $\hat{U}$, $\tilde{U}$ in
front of $\mu_1$, $\Delta_{ij}$ range in
$$[9,7882] \cap \Z\;,\qquad [9,184] \cap \Z\;,$$
respectively. Hence, $U>0$ and none of the expressions $NS_q\Xi_\pm$ vanishes, and $g_{\psi\psi}>0$ on $V_{II}$.

In order to study $g_{\psi\psi}$ on $V_{III}$ ($\rho=\mu_1=0$, $\mu_5>0$) we substitute $\mu_1=\rho^2\hat{\mu}_1$
into $\Xi_\pm$ (the numerator of $g_{\psi\psi}$ is given by
$\Xi_+\Xi_-$). It turns out that $\rho^4$ factors out in each
term $\Xi_\pm$. On the other hand, $\rho^8$ factors in
$\mu_1\mu_4 H_x F$ (the denominator of $g_{\psi\psi}$). We set
$$\breve\Xi_\pm=\lim_{\rho\rightarrow 0^+}\frac{\Xi_\pm}{\rho^4}$$
and repeat the proof for $\breve\Xi_\pm$ as in the case $\rho>0$. We have
\begin{equation*}
U=\frac{1}{4\hat\mu_1^2\mu_3\mu_4\mu_5(1+\hat\mu_1\mu_5)^2}\frac{\hat{U}}{\tilde{U}}\;.
\end{equation*}
Finally, the coefficients in  the polynomials $\hat{U}$,
$\tilde{U}$ in front of the strictly positive functions
$\hat\mu_1$, $\Delta_{ij}$ are again greater than zero and in
$$[9,8714] \cap \Z\;,\qquad [9,184] \cap \Z\;,$$ respectively. Therefore, $g_{\psi\psi}>0$ on $V_{III}$, as expected.

We continue our analysis on $V_{IV}$ ($\rho=\mu_4=0$, $\mu_3>0$). Here,
$\mu_1=\rho^2\hat{\mu}_1$, $\mu_5=\rho^2\hat{\mu}_5$,
$\mu_4=\rho^2\hat{\mu}_4$. We note that
$0<\hat\mu_1<\hat\mu_5<\hat\mu_4$. The calculations
are similar to the calculations for $V_{III}$. A factor
$\rho^{16}$ appears both in the numerator and the denominator
of $g_{\psi\psi}$, thus we define
$$\breve\Xi_\pm=\lim_{\rho\rightarrow 0^+}\frac{\Xi_\pm}{\rho^8}\;,$$
and apply our standard analysis to $\breve\Xi_\pm$.
The formula for $U$ is
\begin{equation*}
U=\frac{\hat\mu_5}{4\hat\mu_1^2\hat\mu_4\mu_2^2\mu_3^3(\hat\mu_1-\hat\mu_5)^2(1+\hat\mu_1\mu_2)^2}\hat{U}\;.
\end{equation*}
In the final step we introduce $\Delta_{51}=\rho^2\hat\Delta_{51}$, $\Delta_{45}=\rho^2\hat\Delta_{45}$ and the coefficients in $\hat{U}$ in front of strictly positive functions $\hat\mu_1$, $\hat\Delta_{51}$, $\hat\Delta_{45}$, $\Delta_{34}$, $\Delta_{23}$ are greater than zero and belong to $[9,5852] \cap \Z$.
Hence, $g_{\psi\psi}>0$ on $V_{IV}$, as desired.

In the case of the rotation axis of the periodic Killing field
$\partial_\psi$ (the set $V_\psi$ where $\rho=\mu_i=0$),
the analysis is slightly more involved. We substitute
$\mu_i=\rho^2\hat\mu_i$ into $\Xi_\pm$. Then, we have verified
that $\Xi_\pm\sim \rho^{15}$ for generic $c_1$ and $q$.
However,  a {\sc Mathematica} calculation reveals that for our
choice of $c_1^2$ and $q$ (the equations \eqref{eq:c1},
\eqref{eq:q}) the leading terms vanish and we have at least
$\Xi_\pm\sim\rho^{16}$. Therefore, we drop the leading terms in
$\Xi_\pm$ and analyse the remaining higher order terms. We
denote them with $\Xi_\pm'$. Next, we set
\begin{equation*}
\breve\Xi_\pm=\lim_{\rho\rightarrow 0^+}\frac{\Xi_\pm'}{\rho^{16}}\;,
\end{equation*}
and apply our standard procedure to $\breve\Xi_\pm$. We have found that
\begin{equation*}
S_{c_1^2}\Delta_\pm=-\frac{4 \hat\mu_1\hat\mu_2^2\hat\mu_4\hat\mu_5^3 (\hat\mu_1 - \hat\mu_2)^2  (\hat\mu_3 -
   \hat\mu_1)^3 (\hat\mu_4 - \hat\mu_1) (\hat\mu_2 -
   \hat\mu_4)^2}{\hat\mu_5-\hat\mu_1}\;,
\end{equation*}
which is strictly negative because $0<\hat\mu_1<\hat\mu_5<\hat\mu_4<\hat\mu_3<\hat\mu_2$. Thus, we have $\Xi_+\Xi_-\sim \rho^{32}$.
The denominator of $g_{\psi\psi}$ (given by $\mu_1\mu_4H_xF$) behaves like $\rho^{30}$ and $g_{\psi\psi}$ vanishes like $\rho^2$.
This, together with positivity of $g_{\psi\psi}$ for $\rho>0$ and continuity of $g_{\psi\psi}$, implies that
$$\lim_{\rho\rightarrow 0^+}\frac{g_{\psi\psi}}{\rho^2}>0$$
on the axis of the rotation of $\partial_\psi$ and the inequality \eqref{eq:sc} holds. It completes the proof of {\it stable causality} of the Black Saturns' d.o.c.

The generalized Weyl coordinates cover also the sets
\begin{eqnarray*}
V_{br}&=&\{\rho=0,\mu_5=0,\mu_4>0\}\cap  V_\mu\;,\\\nonumber
V_{sb}&=&\{\rho=0,\mu_3=0,\mu_2>0\}\cap  V_\mu\;,
\end{eqnarray*}
which have been ignored in our analysis so far. The set $V_{BR}=V_{br}\cup V_4$
corresponds to the event horizon of a black ring and the set $V_{SB}=V_{sb}\cup V_2$ correspond to
the event horizon of a spherical black hole.\footnote{The sets $V_4$, $V_2$ were defined in the footnote on page 3.} 
The closure of $V_{d.o.c.}$ is given by $V_{d.o.c.}\cup V_{BR}\cup V_{SB}$.
Since $V_4\subset V_{IV}$, $V_2\subset V_\psi$, it remains to apply our analysis on $V_{br}$, $V_{sb}$.

On $V_{br}$ ($\rho=\mu_5=0$, $\mu_4>0$) we substitute $\mu_1=\rho^2\hat\mu_1$, $\mu_5=\rho^2\hat\mu_5$ into $\Xi_\pm$. Then, it turns out that $\rho^7$ factors out of $\Xi_\pm$ and $\rho^{14}$ factors out of the denominator of $g_{\psi\psi}$ ($\mu_1\mu_4 H_xF$). We define
\begin{equation*}
\breve\Xi_\pm=\lim_{\rho\rightarrow 0^+}\frac{\Xi_\pm}{\rho^{7}}\;,
\end{equation*}
and calculate the discriminants $\Delta_\pm$ of $NS_q\Xi_\pm$ in respect to $c_2$. Next, we substitute $c_1^2$ into $\Delta_\pm$ and obtain
\begin{equation*}
S_{c_1^2}\Delta_\pm=-w_\pm \frac{\hat\mu_1\hat\mu_5^3\mu_2^4\mu_3^4\mu_4(1+\hat\mu_1\mu_2)^2(1+\hat\mu_1\mu_4)^2(\mu_2-\mu_4)^2}{\hat\mu_5-\hat\mu_1}\;,
\end{equation*}
where the coefficient behind $w_\pm$ is strictly greater than
zero. This time $w_\pm$ are polynomials in $\rho$, $\hat\mu_1$,
$\hat\mu_5$, $\mu_4$, $\mu_3$, $\mu_2$. They become explicitly
positive if written in terms of $\hat\mu_1$, $\hat\Delta_{51}$,
$\Delta_{45}$, $\Delta_{34}$, $\Delta_{23}$ with coefficients
in the range $[3,16] \cap \Z$. Then, $S_{c_1^2}\Delta_\pm<0$
and $g_{\psi\psi}>0$ on $V_{br}$.

The analysis of positivity of $g_{\psi\psi}$ on $V_{sb}$ ($\rho=\mu_3=0$, $\mu_2>0$) mimics
the calculations on $V_{br}$. There are the following changes.
We substitute $\mu_1=\rho^2\hat\mu_1$, $\mu_5=\rho^2\hat\mu_5$,
$\mu_4=\rho^2\hat\mu_4$, $\mu_3=\rho^3\hat\mu_4$ into
$\Xi_\pm$. We set
\begin{equation*}
\breve\Xi_\pm=\lim_{\rho\rightarrow 0^+}\frac{\Xi_\pm}{\rho^{12}}\;,
\end{equation*}
and obtain
\begin{align*}
&S_{c_1^2}\Delta_\pm=\\\nonumber
&-\frac{4\hat\mu_1\hat\mu_4\hat\mu_5^3\mu_2^4(1+\hat\mu_1\mu_2)^2(1+\hat\mu_3\mu_2)(1+\hat\mu_4\mu_2)(1+\hat\mu_5\mu_2)(\hat\mu_3-\hat\mu_1)^3(\hat\mu_4-\hat\mu_1)}{\hat\mu_5-\hat\mu_1}\;,
\end{align*}
which is strictly negative. Therefore, $g_{\psi\psi}>0$ on $V_{sb}$.

In summary, our analysis applied to $V_{br}$, $V_{sb}$
establishes that $g_{\psi\psi}>0$ there.  The strict positivity
of $g_{\psi\psi}$ remains valid on $V_4\subset V_{IV}$,
$V_2\subset V_\psi$, as we know from the analysis of the sets
$V_{IV}$, $V_\psi$. Therefore, $g_{\psi\psi}>0$ holds on both
event horizons $V_{BR}$, $V_{SB}$. This implies that the
inequality \eqref{eq:sc} is satisfied on the event horizons.
They may be included to the Black Saturns' domain of {\it
stable causality} and the Black Saturns are {\it stably causal} on the closure of the d.o.c.

We stress that we have not assigned a particular value to $c_2$
in our calculations so all results hold also for the Black
Saturns with conical singularities on the axes of the rotation
of the periodic Killing fields $\partial_\psi$,
$\partial_\varphi$.  \vspace{0.4cm}

\noindent{\sc Acknowledgements}

We are grateful to Piotr Chru\'sciel and Micha{\l} Eckstein for
useful discussions. Our calculations were carried out with {\sc
Mathematica}.
This research was supported by the Foundation for Polish
Science.

\bibliographystyle{amsplain}
\bibliography{BScausality}
\end{document}

%% file: LeeNewcommand.tex
\newcommand{\jlcax}[1]{}
\newcommand{\eean}{\nonumber\end{eqnarray}}

\newcommand{\kk}[1]{}

\newcommand{\beq}{\begin{equation}}

\newcommand{\FS}       
                  {F}

\newcommand{\HS} 
       {H_{\mbox{\scriptsize volume}}}

{\ptc{this should be removed in the oberwolfach version}}%

\newcommand{\eeal}[1]{\label{#1}\end{eqnarray}}
\newcommand{\bed}{\begin{deqarr}}
\newcommand{\eed}{\end{deqarr}}
\newcommand{\bedl}[1]{\begin{deqarr}\label{#1}}
\newcommand{\eedl}[2]{\arrlabel{#1}\label{#2}\end{deqarr}}

\newcommand{\bel}[1]{\begin{equation}\label{#1}}
\newcommand{\bea}{\begin{eqnarray}}
\newcommand{\bean}{\begin{eqnarray}\nonumber}
\newcommand{\beal}[1]{\begin{eqnarray}\label{#1}}
\newcommand{\eea}{\end{eqnarray}}

\newcommand{\be}{\begin{equation}}
\newcommand{\eeq}{\end{equation}}
\newcommand{\ee}{\end{equation}}
\newcommand{\beqa}{\begin{eqnarray}}
\newcommand{\eeqa}{\end{eqnarray}}
\newcommand{\beqan}{\begin{eqnarray*}}
\newcommand{\eeqan}{\end{eqnarray*}}
\newcommand{\ba}{\begin{array}}
\newcommand{\ea}{\end{array}}

\newcommand{\mnote}[1]
{\protect{\stepcounter{mnotecount}}$^{\mbox{\footnotesize
$
\bullet$\themnotecount}}$ \marginpar{
\raggedright\tiny\em
$\!\!\!\!\!\!\,\bullet$\themnotecount: #1} }

\newcommand{\warn}[1]
{\protect{\stepcounter{mnotecount}}$^{\mbox{\footnotesize
$
\bullet$\themnotecount}}$ \marginpar{
\raggedright\tiny\em
$\!\!\!\!\!\!\,\bullet$\themnotecount: {\bf Warning:} #1} }

\newcommand{\R}{\mathbb R}

\newcommand{\Z}{\mathbb Z}

\newcommand{\ptc}[1]{\mnote{{\bf ptc:}#1}}

\newcommand{\beqar}{\begin{deqarr}}
\newcommand{\eeqar}{\end{deqarr}}

\newcommand{\beaa}{\begin{eqnarray*}}
\newcommand{\eeaa}{\end{eqnarray*}}